\documentclass[superscriptaddress,aps,preprintnumbers,amsmath,amssymb,prd,nofootinbib,preprint]{revtex4-1}
\pdfoutput=1
\usepackage{graphicx}
\usepackage{epstopdf}
\usepackage{dcolumn}
\usepackage{bm}
\usepackage{hyperref}
\usepackage{color}
\usepackage{amsmath}
\usepackage{cancel}
\usepackage{xpatch}

\numberwithin{equation}{section}

\makeatletter
\renewcommand{\p@subsection}{}
\renewcommand{\p@subsubsection}{}
\makeatother

\makeatletter
\def\l@subsubsection#1#2{}
\makeatother

\hypersetup{colorlinks,citecolor= blue,linkcolor= black}

\begin{document}


\def\a{\alpha}
\def\b{\beta}
\def\c{\varepsilon}
\def\d{\delta}
\def\e{\epsilon}
\def\f{\phi}
\def\g{\gamma}
\def\h{\theta}
\def\k{\kappa}
\def\l{\lambda}
\def\m{\mu}
\def\n{\nu}
\def\p{\psi}
\def\q{\partial}
\def\r{\rho}
\def\s{\sigma}
\def\t{\tau}
\def\u{\upsilon}
\def\v{\varphi}
\def\w{\omega}
\def\x{\xi}
\def\y{\eta}
\def\z{\zeta}
\def\D{\Delta}
\def\G{\Gamma}
\def\H{\Theta}
\def\L{\Lambda}
\def\F{\Phi}
\def\P{\Psi}
\def\S{\Sigma}

\def\o{\over}
\def\beq{\begin{align}}
\def\eeq{\end{align}}
\newcommand{\gsim}{ \mathop{}_{\textstyle \sim}^{\textstyle >} }
\newcommand{\lsim}{ \mathop{}_{\textstyle \sim}^{\textstyle <} }
\newcommand{\vev}[1]{ \left\langle {#1} \right\rangle }
\newcommand{\bra}[1]{ \langle {#1} | }
\newcommand{\ket}[1]{ | {#1} \rangle }
\newcommand{\MeV}{\textrm{ MeV}}
\newcommand{\GeV}{\textrm{ GeV}}
\newcommand{\TeV}{\textrm{ TeV}}
\newcommand{\eV}{\textrm{ eV}}
\newcommand{\meV}{\textrm{ meV}}
\newcommand{\1}{\mbox{1}\hspace{-0.25em}\mbox{l}}
\newcommand{\headline}[1]{\noindent{\bf #1}}
\def\diag{\mathop{\rm diag}\nolimits}
\def\Spin{\mathop{\rm Spin}}
\def\SO{\mathop{\rm SO}}
\def\O{\mathop{\rm O}}
\def\SU{\mathop{\rm SU}}
\def\U{\mathop{\rm U}}
\def\Sp{\mathop{\rm Sp}}
\def\SL{\mathop{\rm SL}}
\def\tr{\mathop{\rm tr}}
\def\mpl{M_{\rm Pl}}

\def\IJMP{Int.~J.~Mod.~Phys. }
\def\MPL{Mod.~Phys.~Lett. }
\def\NP{Nucl.~Phys. }
\def\PL{Phys.~Lett. }
\def\PR{Phys.~Rev. }
\def\PRL{Phys.~Rev.~Lett. }
\def\PTP{Prog.~Theor.~Phys. }
\def\ZP{Z.~Phys. }

\def\dd{\mathrm{d}}
\def\ff{\mathrm{f}}
\def\BH{{\rm BH}}
\def\inf{{\rm inf}}
\def\ev{{\rm evap}}
\def\eq{{\rm eq}}
\def\SM{{\rm sm}}
\def\Mpl{M_{\rm Pl}}
\newcommand{\Red}[1]{\textcolor{red}{#1}}
\newcommand{\TL}[1]{\textcolor{blue}{\bf TL: #1}}

\title{
Predictions for Axion Couplings from ALP Cogenesis
}
\preprint{LCTP-20-11}

\author{Raymond T.~Co}
\affiliation{Leinweber Center for Theoretical Physics, Department of Physics, University of Michigan, Ann Arbor, MI 48109, USA}
\author{Lawrence J.~Hall}
\affiliation{Department of Physics, University of California, Berkeley, CA 94720, USA}
\affiliation{Theoretical Physics Group, Lawrence Berkeley National Laboratory, Berkeley, California 94720, USA}
\author{Keisuke Harigaya}
\affiliation{School of Natural Sciences, Institute for Advanced Study, Princeton, NJ 08540, USA}

\begin{abstract}
Adding an axion-like particle (ALP) to the Standard Model, with a field velocity in the early universe, simultaneously explains the observed baryon and dark matter densities. This requires one or more couplings between the ALP and photons, nucleons, and/or electrons that are predicted as functions of the ALP mass. These predictions arise because the ratio of dark matter to baryon densities is independent of the ALP field velocity, allowing a correlation between the ALP mass, $m_a$, and decay constant, $f_a$. The predicted couplings are orders of magnitude larger than those for the QCD axion and for dark matter from the conventional ALP misalignment mechanism. As a result, this scheme, ALP cogenesis, is within reach of future experimental ALP searches from the lab and stellar objects, and for dark matter.
\end{abstract}

\date{\today}

\maketitle

\tableofcontents

\section{Introduction}
A theory of particle interactions in the early universe offers the remarkable prospect that the contents of the universe can be computed. Unfortunately, the Standard Model of particle physics (SM) does not contain a candidate for the dark matter and, when combined with the hot expanding universe, does not yield a satisfactory calculation of either the observed baryon density or the dark energy density. 

Leaving aside the question of dark energy, which could be a cosmological constant environmentally selected on a multiverse~\cite{Weinberg:1987dv}, in this paper we propose that the baryon and dark matter densities can be simultaneously explained by a single new ingredient added to the SM: an Axion-Like Particle (ALP) that possesses an initial field velocity $\dot{\theta}$ and couplings to SM particles. We compute the required ALP couplings to photons, electrons and nucleons and find them to be orders of magnitude larger than for the QCD axion~\cite{Peccei:1977hh,Peccei:1977ur,Weinberg:1977ma,Wilczek:1977pj} and for dark matter from conventional misalignment~\cite{Preskill:1982cy,Abbott:1982af,Dine:1982ah} for the ALP. Proposed experiments probing ALP couplings will test our cogenesis scheme for baryon and dark matter abundances by verifying the correlation between the ALP coupling and mass.

Our mechanism depends on the mass $m_a$ and decay constant $f_a$ of the ALP, together with the comoving charge density associated with the broken $U(1)$ symmetry, $Y_\theta \propto \dot{\theta}$. Dark matter is composed of a condensate of zero momentum ALPs produced by the kinetic misalignment mechanism \cite{Co:2019jts} with an energy density to entropy ratio of 
\begin{align}
    \frac{\rho_a}{s} \simeq 2m_a \, Y_\theta.
\end{align}
On the other hand, the charge asymmetry in the ALP condensate $Y_\theta$ gets transferred via its SM couplings to particle-antiparticle asymmetries of SM particles in the thermal bath, and at temperatures above the electroweak scale the electroweak anomaly converts this to a baryon asymmetry relative to entropy of
\begin{align}
Y_B = &\frac{n_B}{s} =  c_B \, 
\left( \frac{T_{\rm EW}}{f_a} \right)^2 \, Y_\theta
\label{eq:YB}
\end{align}
where $T_{\rm EW} \sim 130$ GeV is the temperature below which the sphaleron process drops out of equilibrium. The constant $c_B$ depends on the ALP coupling to SM particles, and is of order~0.1. The unknown initial charge density $Y_\theta$ drops out of the ratio of the axion to baryon densities, allowing a precise correlation of $f_a$ and $m_a$, and hence the prediction of the ALP couplings as a function of $m_a$.

Our baryogenesis mechanism builds on earlier work.
Baryogenesis from condensation of a scalar field is discussed in the literature. A rotating condensate that carries baryon charge can decay into quarks and produce baryon asymmetry, which is called Affleck-Dine baryogenesis~\cite{Affleck:1984fy}. Spontaneous Baryogenesis~\cite{Cohen:1987vi,Cohen:1988kt} relies on the angular velocity of the condensate that acts as an effective chemical potential for a thermal bath, generating a baryon asymmetry for the quarks using a baryon number violating interaction. Baryogenesis can result from a condensate carrying charge $Q$ other than baryon number~\cite{Chiba:2003vp,Takahashi:2003db}, although they require an interaction that violates both $Q$ and $B$ to be in thermal equilibrium. The baryon number violation by the weak sphaleron process~\cite{Klinkhamer:1984di,Kuzmin:1985mm} is utilized in leptongenesis~\cite{Fukugita:1986hr,Davidson:2008bu}, electroweak baryogenesis~\cite{Kuzmin:1985mm, Cohen:1990it, Cohen:1990py}, and together with the strong sphaleron process in axiogenesis~\cite{Co:2019wyp}.

In Sec.~\ref{sec:ALP_cogenesis}, we analyze the ALP cogenesis mechanism in detail, while providing a precise analytic computation for $c_B$ of Eq.~(\ref{eq:YB}) in Appendix A. In Sec.~\ref{sec:ALP_couplings} we give precise predictions for the ALP coupling to photons, electrons and nucleons, and compare these predictions with reaches of proposed experiments.  The origin of the ALP velocity $\dot{\theta}$ is briefly discussed in Sec.~\ref{sec:initiation}, and conclusions are discussed in Sec.~\ref{sec:summary}.

\section{The ALP Cogenesis Framework}
\label{sec:ALP_cogenesis}
In this section we present the framework that leads to ALP baryogenesis at the weak scale and generates ALP dark matter from kinetic misalignment below the weak scale.
\subsection{The EFT at the Weak Scale}

We take the Effective Field Theory (EFT) at the weak scale to be the Standard Model together with an ALP  that has some non-zero couplings to SM particles.  The SM has a $U(3)^5$ flavor symmetry acting on the left-handed Weyl fields for quarks and leptons, $f_i = q_i, \bar{u}_i. \bar{d}_i, \ell_i, \bar{e}_i$, explicitly broken by the Yukawa interactions
\begin{align}
    {\cal L}_{\rm Yukawa} = y^u_{ij}\; q_i \bar{u}_j H^\dag + y^d_{ij} \; q_i \bar{d}_j H + y^e_{ij} \; \ell_i \bar{e}_j H + h.c.
\end{align}
where $H$ is the Higgs field.

The UV extension of the EFT possesses some global symmetry, $U(1)_P$, that is spontaneously broken by a field $P$, with radial and angular excitation modes $S$ and $\theta$, and a zero temperature vacuum value given by the ALP symmetry breaking scale $f_a$
\begin{align}
\label{eq:P}
    P = \frac{1}{\sqrt{2}}(f_a N_{\rm DW} + S) \; e^{i\theta/N_{\rm DW}}.
\end{align}
$N_{\rm DW}$ is the domain wall number, determined by how $U(1)_P$ is explicitly broken and the resulting ALP potential. In the low energy EFT, the ALP $a = \theta f_a$ is assumed to have a potential
\begin{align}
\label{eq:VALP}
    V(a) = m_a^2f_a^2 \left( 1- {\rm cos}\frac{a}{f_a} \right)
\end{align}
which is periodic in $a/f_a$ with period $2\pi$. The ALP mass is $m_a$. Even if the mass of $S$ is less than the weak scale, it is very weakly coupled to the SM and and the potential and couplings of $S$ are not needed in this paper.

If the UV completion does not involve fermions beyond those of the SM, $U(1)_P$ is a sub-group of the $U(3)^5$ flavor group, but in the presence of additional heavy fermions with SM gauge quantum numbers, $U(1)_P$ may lie partly or wholly outside $U(3)^5$. In the weak scale EFT, the most general set of interactions between the ALP and SM particles up to dimension 5 is
\begin{align}
\label{eq:LALP}
    {\cal L}_\theta = \frac{\partial_\mu a}{f_a} \sum_{f,i,j}c_{f_{ij}} \, f_i^\dag \bar{\sigma}^\mu f_j + \frac{\partial_\mu a}{64\pi^2 f_a} \left( c_Y\, g'^{2}B^{{\mu\nu}} B_{\mu\nu}+ c_W g^{2}\; W^{\mu\nu} W_{\mu\nu} + c_g\; g_3^{2} \, G^{\mu\nu} G_{{\mu\nu}} \right),
\end{align}
where $g', g, g_3$ are gauge couplings and $B_{\mu\nu}, W_{\mu\nu}, G_{\mu\nu}$ are field strengths of the $U(1), SU(2)_L$,  $SU(3)$ SM gauge interactions. Without loss of generality, we work in a basis where $a$ is derivatively coupled and we perform a hypercharge rotation to set the $U(1)_P$ charge of $H$ to zero.

In the simplest theories, couplings $c_{f_{ij}}$ are proportional to the $U(1)_P$ charges of $f_i$. The couplings $c_{Y,W,g}$ denote anomaly coefficients of the shift symmetry on $a$ and are rational numbers. Anomalous field re-definitions imply that $c_Y-c_W$ is not independent of $c_{f_{ij}}$.  In this paper we study theories with a single axion field, which is an ALP rather than the QCD axion studied in Ref.~\cite{Co:2019wyp} with $c_f = 0$, so we insist that $U(1)_P$ has no QCD anomaly and we set $c_g=0$.
Although we need only this EFT for this paper, we present few examples of the UV completion of the ALP coupling in Sec.~\ref{sec:UV}.

The key that allows the ALP to generate a baryon asymmetry and account for the dark matter abundance is its cosmological evolution: $\dot{\theta}$ must be non-zero at the weak scale.  A non-zero $\dot{\theta}$ satisfies the out-of-equilibrium and CP violation conditions for baryogenesis, and implies that the ALP dark matter abundance does not depend on an initial misalignment angle. The baryon and dark matter number densities are both proportional to $\dot{\theta}$, which drops out of the ratio. In this paper, we do not analyze in detail the cosmological evolution of $P$ in various models, as we have done this in elsewhere for both quadratic and quartic potentials~\cite{Co:2019jts}. In general, a relatively flat potential for $S$ is needed, together with a large initial field value, for example from inflation. An important aspect is the need for explicit symmetry breaking of $U(1)_P$ at high temperatures to generate a large initial $\dot{\theta}$, as discussed in Sec.~\ref{sec:initiation}.

\subsection{ALP baryogenesis at the weak scale}

In general, in the early universe a non-zero velocity of the ALP, $\dot{\theta}\neq 0$, produces a baryon asymmetry.  At temperature $T$, if $S(T)$ is small compared with $f_a$, the rotating ALP field contains a charge density of $U(1)_P$
\begin{align}
\label{eq:ntheta}
    n_\theta = \dot{\theta} f_a^2.
\end{align}

At the weak scale, the $B+L$ anomaly of the $SU(2)_L$ gauge interaction is in thermal equilibrium and if the coupling $c_W$ defined in Eq.~(\ref{eq:LALP}) is non-zero, the $U(1)_P$ charge is partially transferred into a $B+L$ asymmetry $n_{B+L}$~\cite{Co:2019wyp}.
Similarly, if the couplings $c_{f_i}$ are order unity, the interactions of the ALP with SM quarks and leptons $f_i$ are in thermal equilibrium at the weak scale. Scatterings via Yukawa couplings allow a sharing of the charge density between the ALP and the quarks and leptons, so that $\dot{\theta}\neq 0$ creates charge asymmetry densities $n_{f_i}=f_i^\dag \bar{\sigma}^0 f_i$ for fermions $f_i$.  
The charge density associated with the shift symmetry of the ALP, conserved up to dilution from expansion, is a linear combination of $n_\theta$ and the fermion number densities
\begin{align}
\label{eq:charge_a_f}
    n_P = n_{\theta} + \sum_{f,i,j} c_{f_{ij}} f^\dag \bar{\sigma}^0 f.
\end{align}
The $q$ and $\ell$ number densities are transferred into a $B+L$ asymmetry via the electroweak sphaleron process.

In the general case, with $c_W$ and $c_{f_{ij}}$ both non-zero, the net result of the ALP and spaleron interactions being in thermal equilibrium is that $n_{B}$ and $n_L$ reach equilibrium values,
\begin{align}
\label{eq:nB}
    n_B = - n_L = c_B \dot{\theta} T^2,
\end{align}
where $c_B$ is a constant given by 
\begin{align}
    c_B=&  - \frac{12}{79} c_W+ \sum_i \left(\frac{18}{79} c_{q_{ii}} - \frac{21}{158}c_{\bar{u}_{ii}} -  \frac{15}{158}c_{\bar{d}_{ii}} + \frac{25}{237} c_{\ell_{ii}}- \frac{11}{237}c_{\bar{e}_{ii}} \right) \\
    \simeq  &
     -0.15 c_W  + 0.68 c_q - 0.40c_{\bar{u}} - 0.28c_{\bar{d}} + 0.32 c_\ell-0.14c_{\bar{e}},  \nonumber
\end{align}
where in the second line we assume flavor universal and diagonal couplings, $c_{f_{ij}} = \delta_{ij} c_f$. 
We derive the coefficient $c_B$ in the Appendix using the picture of the charge transfer from $n_{\theta}$.
The same result can be derived by regarding $\dot{\theta}$ as a background field and the couplings in Eq.~(\ref{eq:LALP}) as effective chemical potentials of the Chern-Simons number and fermion numbers.

As long as $f_a \gg T$, $n_\theta$ almost does not change because of the sharing and remains $\simeq \dot{\theta} f_a^2$~\cite{Co:2019wyp}. This is because it is free-energetically favorable to keep the approximately conserved charge in $n_\theta$ rather than in the asymmetries of particle excitations. For $c_g=0$, since a linear combination of the ALP shift symmetry and fermion numbers remains exact up to the explicit breaking by the ALP potential, the ALP velocity is not damped by the ALP-SM couplings as long as $|\dot{\theta}| \gg m_a$.%
\footnote{Even if $c_g \neq 0$, the damping rate is about $y_u^2 T^3 / f_a^2$ and negligible.}

The sphaleron process ceases to be effective after the electoweak phase transition. In the Standard Model, the temperature below which the sphaleron process is ineffective, $T_{\rm EW}$, is around $130$ GeV~\cite{DOnofrio:2014rug}. Baryon asymmetry is conserved at $T<T_{\rm EW}$,
\begin{align}
Y_B = &\frac{n_B}{s} 
= \left. c_B \frac{\dot{\theta} T^2}{s} \right|_{T = T_{\rm EW}}
=  c_B Y_\theta 
\left( \frac{T_{\rm EW}}{f_a} \right)^2 
\left( \frac{f_a}{f_a(T_{\rm EW})} \right)^2 \\
&= 8.5\times 10^{-11} 
\left(\frac{c_B}{0.1} \right)
\left( \frac{Y_{\theta}}{500} \right) 
\left( \frac{{\rm 10^8 GeV}}{f_a} \right)^2  
\left(\frac{T_{\rm EW}}{130~{\rm GeV}}\right)^2 \left( \frac{f_a}{f_a(T_{\rm EW})} \right)^2 \nonumber
\end{align}
Here we take into account the possibility that the decay constant around the electroweak phase transition, $f_a(T_{\rm EW})$, is different from the present value $f_a$.

In general, the decay constant varies throughout the cosmological evolution of the ALP. The ALP is obtained by a spontaneous breaking of $U(1)_P$ by $P$ of Eq.~(\ref{eq:P}).  The charge $n_\theta$ is given by
\begin{align}
    n_{\theta}= & \frac{1}{N_{\rm DW}}\left(i\dot{P^*}P- i\dot{P}P^*\right)= \dot{\theta}\left(f_a + \frac{S}{N_{\rm DW}}\right)^2
\end{align}
generalizing Eq.~(\ref{eq:ntheta}).
In the early universe $S$ is not necessarily at the minimum $S=0$, leading to the decay constant different from the present one,
\begin{align}
f_{a}(T) = f_a + \frac{S(T)}{N_{\rm DW}}.
\end{align}
In fact, in the mechanism generating the ALP velocity discussed in Sec.~\ref{sec:initiation}, $S$ may be larger than $f_a N_{\rm DW}$ in the early universe even around the electroweak phase transition.

\subsection{ALP Dark Matter from kinetic misalignment}

In the conventional misalignment mechanism, the ALP field is stuck at a field value $a_i$ for $H \gg m_a$, and begins to oscillate in the potemntial of Eq.~(\ref{eq:VALP}) when $3H \simeq m_a$. The oscillation behaves as matter, and the resultant energy density of the oscillation $\rho_a$ is
\begin{align}
    \frac{\rho_a}{s} & = 
    \left( \frac{\pi^2 g_*}{10} \right)^{ \scalebox{1.01}{$\frac{3}{4}$}} 
    \left( \frac{45}{2 \pi^2 g_*} \right)
    \frac{ m_a^{1/2} f_a^2 \theta_i^2}{M_{\rm Pl}^{3/2}} 
    \simeq 0.4 \eV \, \theta_i^2
    \left( \frac{m_a}{4 \meV} \right)^{\scalebox{1.01}{$\frac{1}{2}$}}
    \left( \frac{f_a}{10^{12} \GeV} \right)^2
\end{align}
where $g_*$ is the relativistic degrees of freedom at the onset of the oscillation.
Here we normalize the energy density by the entropy density $s$, since after the beginning of the oscillation both $\rho_a$ and $s$ decrease in proportion to $R^{-3}$, where $R$ is the scale factor of the universe. For simplicity, we assumed that the potential is temperature-independent. The observed dark matter abundance is $\rho_{\rm DM} / s \simeq 0.44 \eV$.

The picture may be altered if the ALP has a non-zero initial kinetic energy as proposed in Ref.~\cite{Co:2019jts}. Suppose that the ALP (nearly) coherently evolves, $\dot{\theta} \neq 0$. If the kinetic energy is larger than the potential energy when $H\sim m_a$, the ALP continues to move in the same direction, repeatedly running over the potential barriers. The ALP begins oscillations about the minimum of the potential when the kinetic energy becomes smaller than the potential barrier. The beginning of the oscillation is delayed in comparison with the conventional misalignment mechanism, enhancing the ALP energy density. We named this scenario the kinetic misalignment mechanism.

Let us estimate the ALP energy density.
We parameterize the kinetic energy by
\begin{align}
\label{eq:n_Y_theta}
    n_\theta \equiv \dot{\theta} f_a^2,~~Y_{\theta} \equiv \frac{n_{\theta}}{s},
\end{align}
where $s$ is the entropy density of the universe. Once we understand the ALP as the angular direction, $n_\theta$ is then the angular momentum and a charge density associated with the approximate shift symmetry of the ALP, $\theta \rightarrow \theta + \alpha$. When the kinetic energy is much larger than the potential energy, $n_\theta$ is conserved up to the cosmic expansion $n_\theta \propto R^{-3}$ and thus the yield $Y_\theta$ remains constant.

The kinetic energy $\dot{\theta}^2 f_a^2/2$ becomes comparable to the potential barrier $2m_a^2 f_a^2$ when $\dot{\theta} = 2m_a$. The entropy density $s$ at this point is $2m_af_a^2/Y_{\theta}$. The ALP begins oscillation with an initial number density $\simeq 2 m_a f_a^2$. The number density $n_a$ of the oscillating ALP is
\begin{align}
    Y_a \equiv \frac{n_a}{s} = C\frac{2m_af_a^2}{2 m_a f_a^2/Y_\theta} = C Y_\theta.
\end{align}
Here $C$ is a numerical factor taking into account the deviation from the analytical estimation due to the anharmonicity around the hilltop of the potential. The numerical computation performed in Ref.~\cite{Co:2019jts} finds that $C\simeq 2$.
Note that the estimation of $Y_a$ is valid even if the ALP potential changes in time as long as the change is adiabatic so that the number density of the oscillation is conserved.
The energy of density of the ALP oscillation by the kinetic misalignment mechanism is
\begin{align}
    \frac{\rho_a}{s} = m_a Y_a \simeq 2m_a Y_\theta = 0.4~{\rm eV} \left( \frac{m_a}{\rm meV} \right) \left( \frac{Y_\theta}{400} \right).
\end{align}

\subsection{ALP Cogenesis: baryon asymmetry and dark matter}
Baryon asymmetry and dark matter density are mainly determined by three parameters: $m_a$, $f_a$ and $Y_{\theta}$. After requiring the baryon and dark matter abundance to be the observed values~\cite{Aghanim:2018eyx},
\begin{align}
    Y_B^{\rm obs} \simeq 8.7\times 10^{-11}, \ \ \ \frac{\rho_{\rm DM}^{\rm obs}}{s} \simeq 0.44~{\rm eV},
\end{align}
we can predict $f_a$ as a function of $m_a$,
\begin{align}
\label{eq:fa_prediction}
    f_a = 2\times 10^9~{\rm GeV} 
    \left( \frac{f_a}{f_a(T_{\rm EW})} \right)
    \left(\frac{c_B}{0.1}\right)^{{ \scalebox{1.01}{$\frac{1}{2}$} }} 
    \left( \frac{1~\mu{\rm eV}}{m_a} \right)^{ \scalebox{1.01}{$\frac{1}{2}$} } 
    \left( \frac{T_{\rm EW}}{130~{\rm GeV}} \right) .
\end{align}
Assuming that the electroweak phase transition is the standard model-like and $f_a(T_{\rm EW}) = f_a$, the decay constant is uniquely predicted. It is possible that $f_a(T_{\rm EW})/f_a > 1$, which reduces the prediction on $f_a$. This predicted value of $f_a$ from ALP cogenesis is typically much smaller than that from the QCD axion
\begin{align}
\label{eq:fa_QCD}
    f_a = 6 \times 10^{12} \GeV \left( \frac{\mu{\rm eV}}{m_a} \right) ,
\end{align}
and from ALP dark matter with the conventional misalignment mechanism
\begin{align}
\label{eq:fa_MM}
    f_a =       10^{13} \GeV \ \theta_i^{-1}\left(\frac{\mu{\rm eV}}{m_a}\right)^{\scalebox{1.1}{$\frac{1}{4}$} } .
\end{align}
The smaller decay constant means larger couplings of the ALP with standard model particles.
We discuss how the predicted value is probed by ALP search.

We assume that the oscillation of the ALP begins after the electroweak phase transition.
For a temperature independent ALP potential, this assumption is consistent if 
\begin{align}
    m_a < 3~{\rm keV} 
    \left( \frac{0.1}{c_B} \right)
    \left( \frac{T_{\rm EW}}{130~{\rm GeV}} \right)
    \left( \frac{f_a(T_{\rm EW})}{f_a} \right)^2.
\end{align}
The constraint becomes weaker if the ALP potential is suppressed at high temperatures.

\section{ALP couplings}
\label{sec:ALP_couplings}

Our mechanism for baryogenesis requires ALP couplings to standard model particles. This should be contrasted with ALP dark matter from the misalignment mechanism, where no couplings with SM particles are required. In this section, we discuss how the couplings arise from UV completions, and how predicted couplings can be probed by future experiments.

\subsection{UV completions}
\label{sec:UV}

We discuss a few UV completions of the ALP couplings realizing various hierarchies of $c_{W,Y}$ and $c_f$. For $c_f$, we mainly introduce a model which gives non-zero $c_\ell$, but a generalization to other $c_f$ is straightforward.

\subsubsection{$c_{W,Y} = \mathcal{O}(1)$, $|c_f| \ll 1$}
Non-zero $c_{W,Y}$ arises from the anomaly of $U(1)_P$ symmetry. The simplest example is a model of heavy $U(1)_P$-charged $SU(2)_L$ doublet fermions $L$ and $\bar{L}$ obtaining mass from $P$,
\begin{align}
    {\cal L} = \lambda P L \bar{L} \rightarrow c_W = \frac{1}{N_{\rm DW}}.
\end{align}

Even if $c_f$ vanish at tree-level, they are generated from $c_{W,Y}$ by one-loop radiative corrections. From $c_W$, non-zero $c_{q,\ell}$ are generated as~\cite{Bauer:2017ris}
\begin{align}
\label{eq:cf_loop}
    c_{q,\ell} \simeq \frac{9}{2} \left(\frac{\alpha_2}{4\pi}\right)^2 {\rm ln}\frac{\Lambda}{v_{\rm EW}} \times c_W \simeq 7\times 10^{-4} \left( \frac{{\rm ln}(\Lambda/v_{\rm EW})}{20} \right) \times c_W,
\end{align}
where $\Lambda$ is the scale where the coupling $c_W$ is generated.

\subsubsection{$c_{W,Y} = \mathcal{O}(1)$, $c_f = \mathcal{O}(1)$}
If the SM fermions have a non-zero $U(1)_P$ charge, $\mathcal{O}(1)$ $c_f$ arises. 
Let us consider a heavy fermion $L$ which has the same gauge quantum number as $\ell$, and its Dirac partner $\bar{L}$. Choosing $U(1)_P$ charges $\ell(-1)$, $\bar{e}(0)$, $L(0)$ and $\bar{L}(0)$, the renormalizable couplings are
\begin{align}
\label{eq:cW1cf1}
    {\cal L}= (m L + \lambda P \ell) \bar{L} + y L\bar{e}H.
\end{align}
Assuming $m \gg \lambda \vev{P}$, after integrating out $L\bar{L}$, we obtain
\begin{align}
    {\cal L} = - y \lambda \frac{f_a N_{\rm DW}}{\sqrt{2} m} e^{i \theta/N_{\rm DW}} \ell \bar{e}H - \frac{\lambda^2 f_a^2 N_{\rm DW}}{2 m^2} \partial_\mu \theta \ell^\dag \bar{\sigma}^\mu \ell.
\end{align}
After eliminating $\theta$ from the Yukawa coupling by the rotation of $\ell$, we obtain
\begin{align}
    c_W = c_Y = \frac{1}{N_{\rm DW}},~c_\ell = \frac{1}{N_{\rm DW}} - \frac{\lambda^2 f_a^2 N_{\rm DW}}{2m^2}.
\end{align}
The coupling $c_\ell$ is smaller than $1/N_{\rm DW}$ since the SM $\ell$ is an admixture of $\ell(-1)$ and $L(0)$.
The structure in Eq.~(\ref{eq:cW1cf1}) is nothing but that of the Froggatt-Nielsen model of flavor~\cite{Froggatt:1978nt}.

\subsubsection{$c_{W,Y} = 0$, $c_f = \mathcal{O}(1)$}
If the SM fermions have non-zero $U(1)_P$ charges but the $U(1)_P$ symmetry does not have quantum anomaly, $c_{W,Y}=0$ while $c_f = \mathcal{O}(1)$. Choosing $U(1)_P$ charges $\ell(-1)$, $\bar{e}$, $L(0)$ and $\bar{L}(1)$, the renormalizable couplings are
\begin{align}
\label{eq:cW0cf1}
    {\cal L}= ( \lambda P^\dag L + m \ell) \bar{L} + y L\bar{e}H.
\end{align}
Assuming $\lambda\vev{P} \gg m$, after integrating out $L\bar{L}$, we obtain
\begin{align}
    c_\ell = \frac{1}{N_{\rm DW}} -  \frac{2 m^2}{\lambda^2 f_a^2 N_{\rm DW}}.
\end{align}

\subsubsection{$c_{W,Y} = 0$, $|c_f| \ll 1$}
If the SM fermions do not have $U(1)_P$ charges and the $U(1)_P$ symmetry does not have quantum anomaly, the ALP couplings may be suppressed. Non-zero $c_f$ arises from mixing of the SM fermions with $U(1)_P$ charged heavy fermions. In the model in Eq.~(\ref{eq:cW0cf1}), if $\lambda\vev{P} \ll m$, the SM fermions is mainly $L$ rather than $\ell$. Through the mixing, we obtain
\begin{align}
    c_\ell =  \frac{\lambda^2 f_a^2 N_{\rm DW}}{2m^2}.
\end{align}

\subsection{Experimental probes of ALP couplings}

We discuss how the predicted ALP couplings can be probed. We consider cases where $c_{W,Y} = \mathcal{O}(1)$ or $c_{f_1} = \mathcal{O}(1)$. 

\subsubsection{Photons}

When the baryon asymmetry is produced by the weak anomaly of the ALP shift symmetry, the ALP is predicted to couple with photons,
\begin{align}
{\cal L} &= -\left(\frac{g_{a\gamma\gamma}}{4} \right) a \,  \epsilon^{\mu\nu\rho\sigma}F_{\mu\nu}F_{\rho\sigma} ,
\end{align}
with a strength predicted by the baryon asymmetry and dark matter abundance to be
\begin{align}
\left|g_{a\gamma \gamma} \right| 
 &= \frac{ \alpha |c_\gamma| }{2 \pi f_a}  
 \simeq 1.8 \times 10^{-11} \GeV^{-1} |c_\gamma| \left( \frac{f_a(T_{\rm EW})}{f_a} \right) \left( \frac{0.1}{c_B} \right)^{ \scalebox{1.01}{$\frac{1}{2}$}} \left( \frac{m_a}{\rm meV} \right)^{ \scalebox{1.01}{$\frac{1}{2}$}} \left( \frac{130 \GeV}{T_{\rm EW}} \right) 
\end{align}
and $c_\gamma \equiv c_W + c_Y$. For later convenience, we define
\begin{align}
\label{eq:cagg}
c_{a\gamma\gamma} \equiv |c_\gamma| \left( \frac{f_a(T_{\rm EW})}{f_a} \right) \left(\frac{0.1}{c_B}\right)^{ \scalebox{1.01}{$\frac{1}{2}$}}.
\end{align}
A canonical value of $c_{a\gamma\gamma}$ is $\mathcal{O}(1)$, while larger values are possible if $f_a(T_{\rm EW}/f_a) \gg 1$ or $|c_\gamma| \gg 1$.

\begin{figure}[t]
	\includegraphics[width=0.65\linewidth]{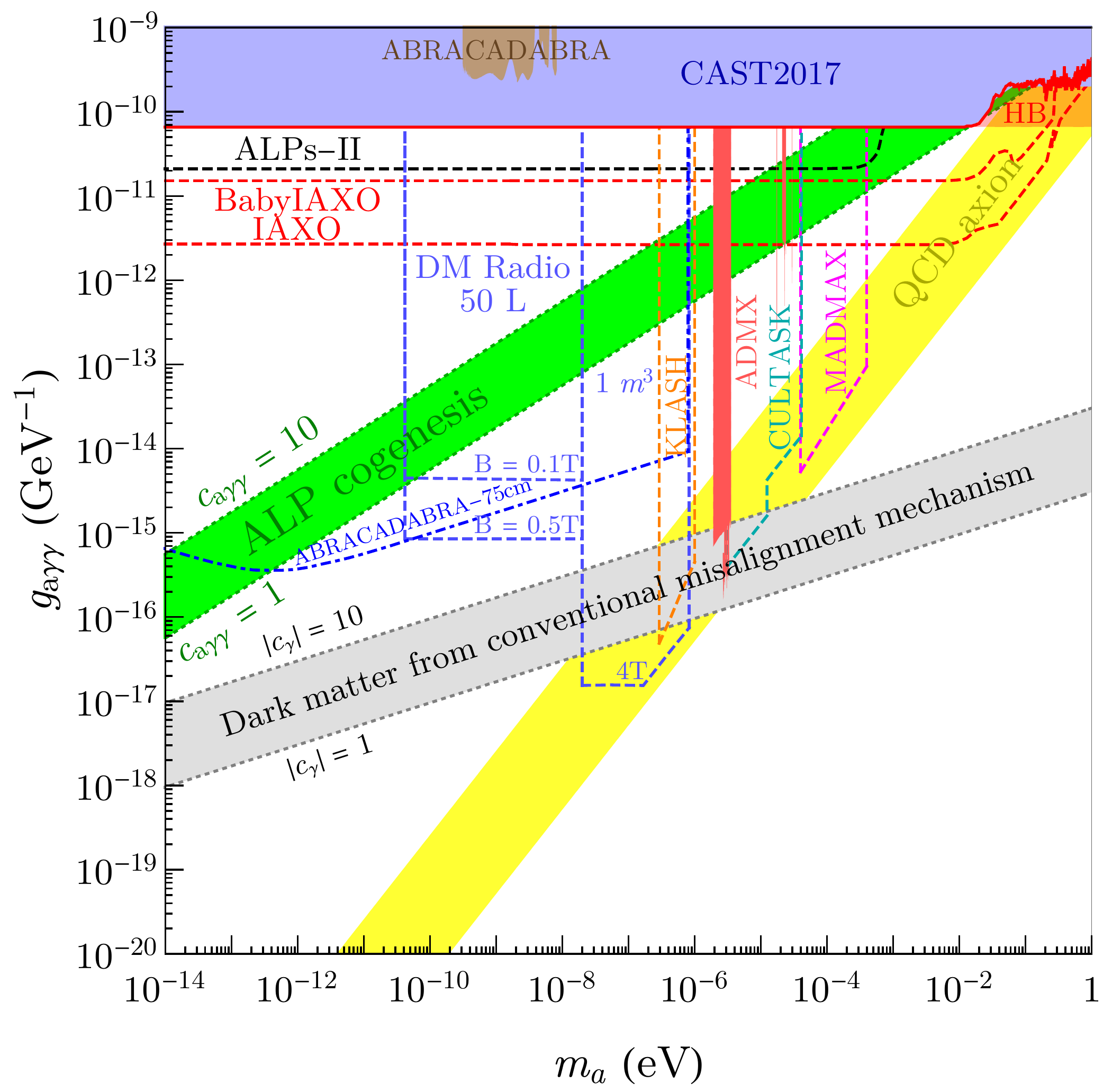}
	\caption{The prediction for the ALP-photon coupling $g_{a\gamma\gamma}$ is shown by the green band, for $c_{a\gamma\gamma} = 1 - 10$. The predictions for the QCD axion and for ALP dark matter from the conventional misalignment mechanism are shown in the yellow and gray bands. Other shaded regions denote the existing experimental constraints, while various lines show the sensitivity of future experiments.}
	\label{fig:photon}	
\end{figure}

Figure~\ref{fig:photon} shows the prediction for the ALP-photon coupling, existing constraints, and the sensitivity of future experiments. The green band shows the prediction from ALP cogenesis as given in Eq.~(\ref{eq:fa_prediction}). The gray band is the usual prediction of ALP dark matter from the conventional misalignment misalignment given in Eq.~(\ref{eq:fa_MM}) with $\theta_i = 1$. The widths of these bands reflect the uncertainty due to the model-dependent constants $c_{a\gamma\gamma}, |c_\gamma|$, respectively, which we vary between 1 and 10. The yellow band corresponds to the QCD axion, defined here with the coefficients of the electromagnetic and strong anomalies, $E$ and $N$, varying over ranges of viable models~\cite{DiLuzio:2020wdo} with $5/3 \le E/N \le 44/3$.

In Fig.~\ref{fig:photon}, we show the regions excluded by CAST~\cite{Anastassopoulos:2017ftl} in blue shading, horizontal branch (HB) stars~\cite{Ayala:2014pea} in red shading, ADMX~\cite{Asztalos:2009yp, Du:2018uak, Boutan:2018uoc, Braine:2019fqb} in red shading, and ABRACADABRA~\cite{Kahn:2016aff, Ouellet:2018beu, Ouellet:2019tlz} in brown shading. The prospects for future proposed and planned experiments are shown individually by the red dashed curves for BabyIAXO and IAXO~\cite{Armengaud:2019uso}, black dashed for ALPs-II~\cite{Bahre:2013ywa}, blue dot-dashed lines for ABRACADABRA~\cite{Kahn:2016aff, Ouellet:2018beu, Ouellet:2019tlz} with a broadband search, blue dashed lines for DM Radio-50L and DM Radio-m$^3$~\cite{Chaudhuri:2014dla, Silva-Feaver:2016qhh, Chaudhuri:2019ntz}, which has merged with ABRACADABRA, for a resonant search\footnote{The planned resonant searches could in principle scan to lower masses and probe more of the ALP cogenesis parameter space.}, cyan dashed lines for CULTASK~\cite{Semertzidis:2019gkj, Lee:2020cfj}, magenta lines for MADMAX~\cite{Brun:2019lyf}, and orange dashed lines for KLASH~\cite{Alesini:2019nzq}. Remarkably, the proposed and planned experiments can probe ALP cogenesis in a wide range of the axion mass.

\subsubsection{Nucleons}
The couplings $c_q$ and $c_{\bar{u},\bar{d}}$ lead to ALP-nucleon interactions, which can be decomposed into couplings of $\partial^\mu a/f_a$ to vector and axial vector currents. For flavor preserving ALP couplings, at an energy scale much below the electroweak scale, only the couplings to the axial vector current are relevant, since the couplings to the vector current can be removed by an ALP-dependent rotation of the quarks. We do not discuss possible signals from flavor violating couplings in this paper.  The couplings to the axial vector current of quarks are given by
\begin{align}
    {\cal L} = C_u \, \frac{\partial_\mu a}{2 f_a} \, \bar{u}\gamma^\mu \gamma^5 u + C_d \, \frac{\partial_\mu a}{2 f_a} \, \bar{d}\gamma^\mu \gamma^5 d, \nonumber \\
    C_u = - (c_q + c_{\bar{u}}),~C_d = - (c_q + c_{\bar{d}}).
\end{align}
The couplings to the axial vector currents of protons and neutrons are
\begin{align}
    {\cal L} & = g_{app} \times \frac{\partial_\mu a}{2 m_p} \, \bar{p} \gamma^\mu \gamma^5 p + g_{ann} \times \frac{\partial_\mu a}{2 m_n} \, \bar{n} \gamma^\mu \gamma^5 n,
\end{align}
where the nucleon couplings are dependant on the quark couplings~\cite{diCortona:2015ldu}
\begin{align}
    g_{aNN} = C_N \frac{m_N}{f_a} \simeq \left( 0.88(3) C_u - 0.39(2) C_d \right) \frac{m_N}{f_a} ,
\end{align}
where $N=p$ or $n$.
Requiring that both the observed baryon asymmetry and dark matter abundance originate from the $U(1)_P$ charge asymmetry yields the prediction 
\begin{align}
\left|g_{aNN} \right| 
 &= |C_N| \frac{m_N}{f_a}
 \simeq 1.4 \times 10^{-11} |C_N| \left( \frac{f_a(T_{\rm EW})}{f_a} \right) \left( \frac{0.1}{c_B} \right)^{ \scalebox{1.01}{$\frac{1}{2}$}} \left( \frac{m_a}{\rm neV} \right)^{ \scalebox{1.01}{$\frac{1}{2}$}} \left( \frac{130 \GeV}{T_{\rm EW}} \right) .
\end{align}
We define
\begin{align}
\label{eq:caNN}
c_{aNN} \equiv |C_N| \left( \frac{f_a(T_{\rm EW})}{f_a} \right) \left(\frac{0.1}{c_B}\right)^{ \scalebox{1.01}{$\frac{1}{2}$}} ,
\end{align}
and note that $c_{aNN}$ can be order unity but depends on $C_N$ and $f_a(T_{\rm EW}) / f_a$.

\begin{figure}[t]
	\includegraphics[width=0.65\linewidth]{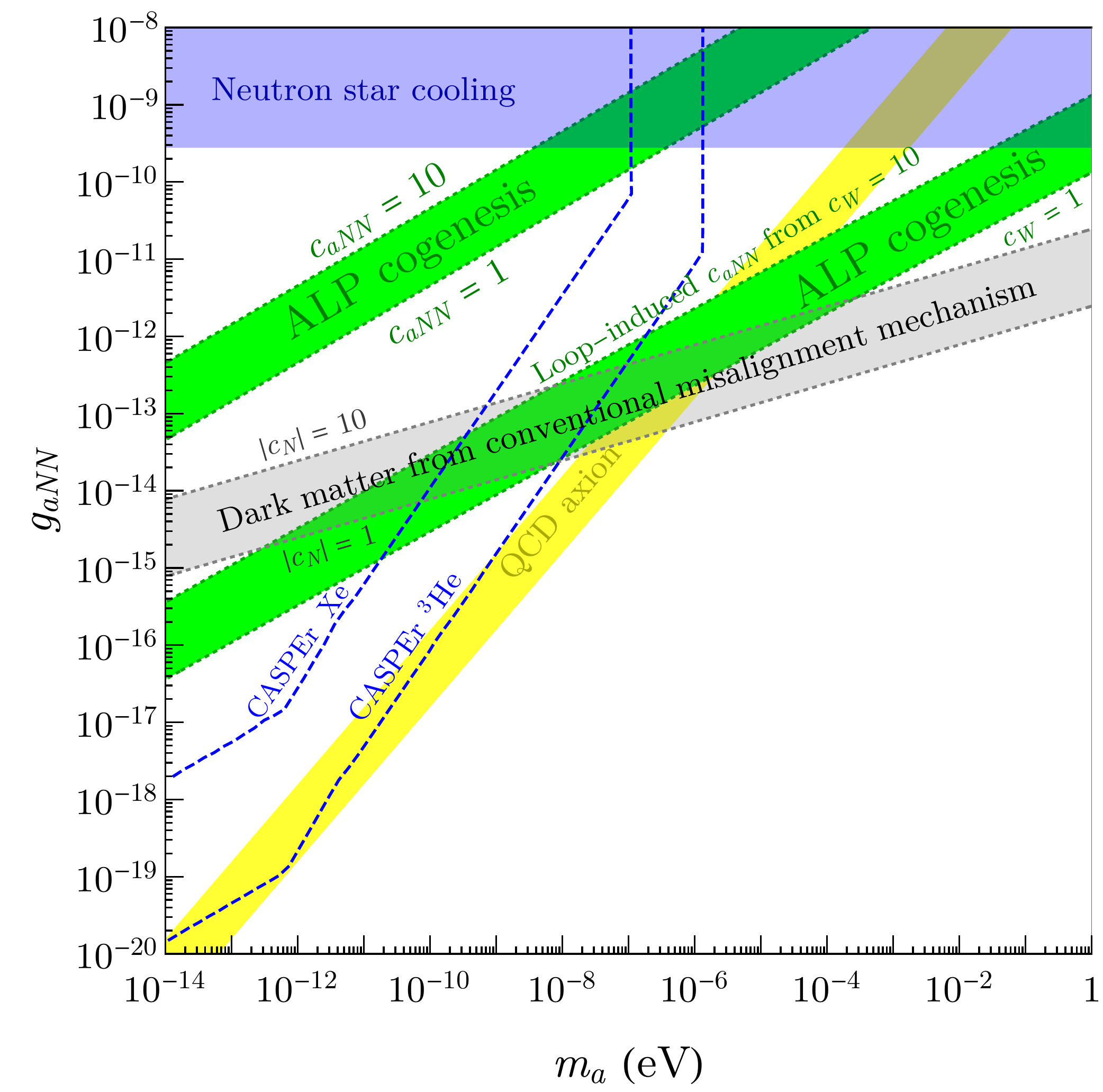}
	\caption{The prediction for the ALP-nucleon coupling $g_{aNN}$ is shown for $c_{aNN} = 1 - 10$ in the upper green band and for loop-induced $c_{aNN}$ from $c_W = 1-10$ in the lower green band. The predictions for the QCD axion and for ALP dark matter from the conventional misalignment mechanism are shown in the yellow and gray bands. The blue shaded region shows the constraint from neutron star cooling, while the blue dashed lines show the sensitivity of the planned experiment CASPEr.}
	\label{fig:nucleon}	
\end{figure}

Fig.~\ref{fig:nucleon} shows the prediction for the ALP-nucleon coupling. The two green bands show the prediction of ALP cogenesis given in Eq.~(\ref{eq:fa_prediction}), in which we vary $c_{aNN} = 1-10$ for the upper band while the lower band is for a much smaller $c_{aNN}$ induced according to Eq.~(\ref{eq:cf_loop}) by $c_W = 1-10$. The gray band shows the usual prediction of ALP dark matter from the conventional misalignment misalignment given in Eq.~(\ref{eq:fa_MM}) with $\theta_i = 1$. The yellow band corresponds to the QCD axion as in Eq.~(\ref{eq:fa_QCD}). The width of these bands reflects the uncertainty in the model-dependent constants $c_{aNN}, |C_N|$, which we vary between 1 and 10. The blue shaded region shows the constraint from neutron star cooling~\cite{Beznogov:2018fda}. The blue dashed lines show the sensitivity for CASPEr~\cite{Budker:2013hfa}.
One can see that CASPEr can probe the ALP cogenesis region with $c_{aNN}$ order unity down to very low ALP masses, and even the loop-suppressed coupling if $m_a \lesssim 10^{-7}$ eV complementing the search using the ALP-photon coupling.

\subsubsection{Electrons}
The couplings $c_\ell$ and $c_{\bar{e}}$ provide interactions of the ALP with electrons. The interaction with the axial current is given by
\begin{align}
    {\cal L} =g_{aee} \times \frac{\partial_\mu a}{2 m_e} \bar{e} \gamma^\mu \gamma^5 e,~~~
    g_{aee}= - (c_\ell + c_{\bar{e}} ) \frac{m_e}{f_a} \equiv c_e \frac{m_e}{f_a}.
\end{align}
ALP cogenesis then predicts
\begin{align}
\left|g_{aee} \right| 
 &= |c_e| \frac{m_e}{f_a}
 \simeq 7.8 \times 10^{-15} |c_e| \left( \frac{f_a(T_{\rm EW})}{f_a} \right) \left( \frac{0.1}{c_B} \right)^{ \scalebox{1.01}{$\frac{1}{2}$}} \left( \frac{m_a}{\rm neV} \right)^{ \scalebox{1.01}{$\frac{1}{2}$}} \left( \frac{130 \GeV}{T_{\rm EW}} \right) ,
\end{align}
and we define 
\begin{align}
\label{eq:caNN}
c_{aee} \equiv |c_e| \left( \frac{f_a(T_{\rm EW})}{f_a} \right) \left(\frac{0.1}{c_B}\right)^{ \scalebox{1.01}{$\frac{1}{2}$}} .
\end{align}

\begin{figure}[t]
	\includegraphics[width=0.65\linewidth]{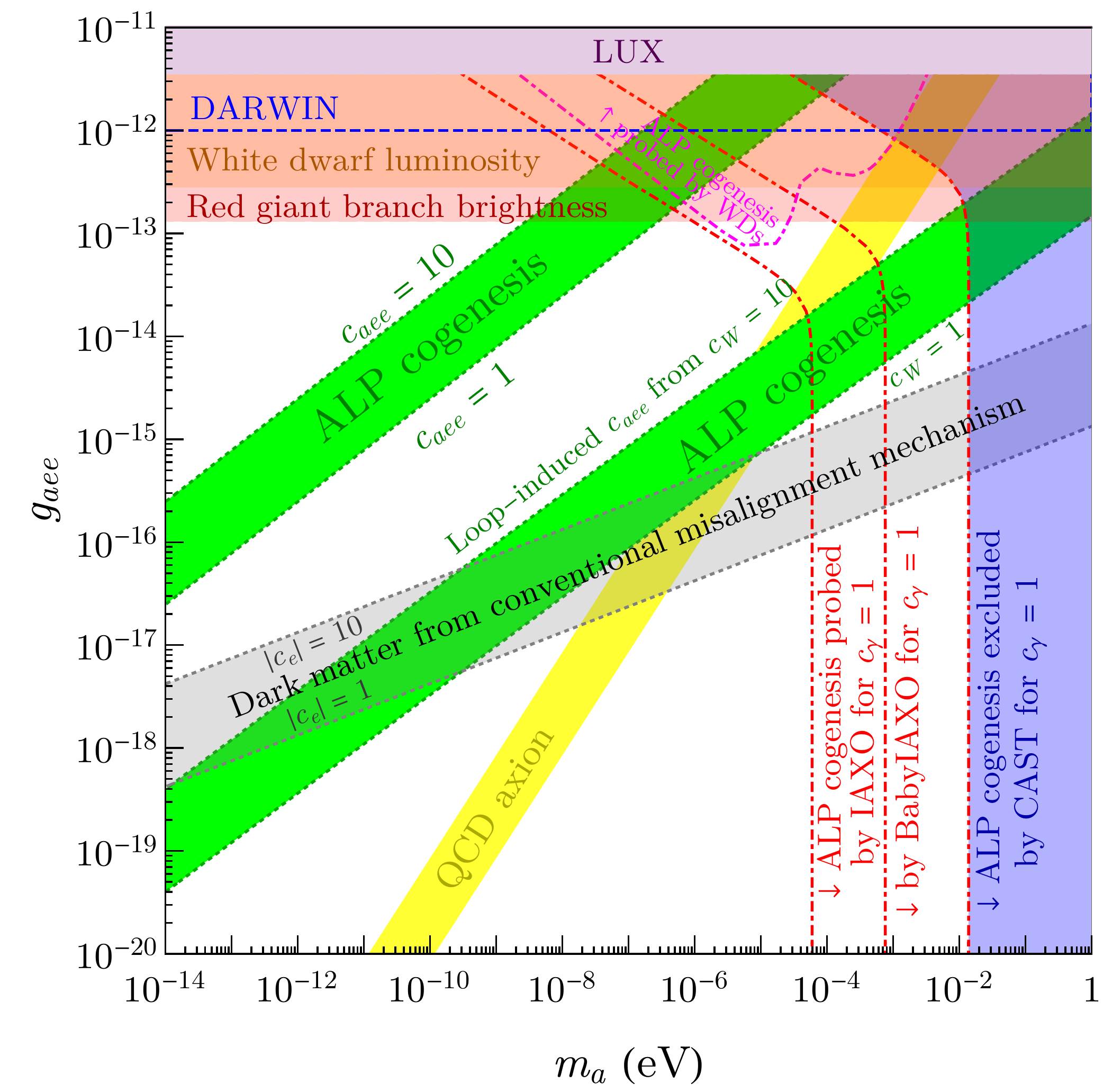}
	\caption{The prediction on the coupling between an ALP and electrons $g_{aee}$ is shown for $c_{aee} = 1 - 10$ in the upper green band and for loop-induced $c_{aee}$ from $c_W = 1-10$ in the lower green band. The prediction of the QCD axions and of ALP dark matter from the conventional misalignment mechanism are shown in the yellow and gray bands. Other shaded regions denote the existing experimental constraints. Various curves show the sensitivity of future experiments, whereas the dot-dashed curves assume ALP cogenesis and $c_{a\gamma\gamma} = 1$.}
	\label{fig:electron}	
\end{figure}  

Fig.~\ref{fig:electron} shows the prediction for the ALP-electron coupling. The two green bands show the prediction of ALP cogenesis given in Eq.~(\ref{eq:fa_prediction}). We vary $c_{aee} = 1-10$ for the upper band, while the lower band is for a much smaller $c_{aee}$ induced  according to Eq.~(\ref{eq:cf_loop}) by $c_W = 1-10$. The gray band shows the usual prediction of ALP dark matter from the conventional misalignment misalignment given in Eq.~(\ref{eq:fa_MM}) with $\theta_i = 1$. The yellow band corresponds to the QCD axion as in Eq.~(\ref{eq:fa_QCD}). The widths of these bands reflect the uncertainty of the model-dependent constant $c_{aee}$, which we vary between $c_{aee} = 1-10$. 

In Fig.~\ref{fig:electron}, we show the regions excluded by the search for solar axions using the underground dark matter direct detection experiment LUX~\cite{Akerib:2017uem} in purple shading, by the luminosity function of white dwarfs~\cite{Bertolami:2014wua} in orange shading, and by the brightness of the red-giant branch~\cite{Viaux:2013lha} in red shading. The future sensitivity of DARWIN~\cite{Aalbers:2016jon} will improve the bound on  $g_{aee}$ from solar axions marked by the blue dashed line. The axion helioscopes constrain the $g_{aee}$-$g_{a\gamma\gamma}$ parameter space, which however can be translated to a limit on $g_{aee}$ when one assumes a value of $c_{a\gamma\gamma}$ and the predicted value of $f_a$ from ALP cogenesis in Eq.~(\ref{eq:fa_prediction}). In this manner, we use $c_{a\gamma\gamma} = 1$ and show the current limit from CAST~\cite{Barth:2013sma} in blue shading and the future prospect of BabyIAXO and IAXO~\cite{Armengaud:2019uso} by red dot-dashed curves. Similarly, a potential reach in $g_{aee}$ is shown by the magenta dot-dashed curve obtained from a future sensitivity on $g_{aee} \times g_{a\gamma\gamma}$ using dedicated $X$-ray observations of the white dwarfs (WDs)~\cite{Dessert:2019sgw} with XMM-Newton~\cite{Brunner:2007kv}.

\subsubsection{ALP gravitational and self-interactions}

Gravitational interactions of ALPs provide a model-independent test. For example, rapidly spinning black holes can release energy and angular momentum via the superradiance mechanism, forming a cloud of ALPs around the black holes. ALP masses between $10^{-13}-10^{-11}$ eV and $10^{-17}-10^{-16}$ eV are excluded for $f_a > \mathcal{O}(10^{14}) \GeV$ and $f_a > \mathcal{O}(10^{16})  \GeV$ from stellar and supermassive black holes respectively~\cite{Arvanitaki:2009fg, Arvanitaki:2010sy, Arvanitaki:2014wva}. However, the values of $f_a$ predicted by ALP cogenesis in Eq.~(\ref{eq:fa_prediction}) are significantly smaller than the constraint and hence axion self-interactions prevent efficient superradiance. Small values of $f_a$ can however lead to a series of bosenova events and potentially produce gravitational wave signals~\cite{Arvanitaki:2014wva}.

\section{Initiation of non-zero ALP velocity}
\label{sec:initiation}

ALP-genesis requires a large charge asymmetry,
\begin{align}
    Y_\theta = 5\times 10^4 \
    \left( \frac{f_a(T_{\rm EW})}{f_a} \right)^2
    \left( \frac{0.1}{c_B} \right) 
    \left(\frac{f_a}{10^9~{\rm GeV}}\right)^2 
    \left(\frac{130~{\rm GeV}}{T_{\rm EW}}\right)^2
\end{align}
which can be obtained by dynamics similar to the Affleck-Dine mechanism~\cite{Affleck:1984fy}. The $U(1)_P$ symmetry may be explicitly broken by a higher-dimensional operator, $V_{PQ}(P) \sim P^n$. If $S$ takes a large initial field value $S_i$ in the early universe, the explicit symmetry breaking is effective and drives the angular motion of $P$. The resultant asymmetry is
\begin{align}
    Y_\theta = & 2\times 10^4 \ \epsilon 
    \left( \frac{S_i}{10^{16}~{\rm GeV}} \right)^2 
    \left( \frac{\rm GeV}{m_{S,i}} \right)^{ \scalebox{1.01}{$\frac{1}{2}$}} 
    \left(\frac{106.75}{g_*} \right)^{ \scalebox{1.01}{$\frac{1}{4}$}},\nonumber\\
    m_{S,i}^2 \equiv & \left.\frac{\partial^2 V}{\partial S^2}\right|_{P=P_i},~~ \epsilon \simeq   \left.\frac{\partial V / \partial \theta}{S \partial V/\partial S}\right|_{P = P_i}.
\end{align}
The large charge asymmetry requires that the initial field value is large while $m_{S,i}$ is small. This requires a flat potential for $S$, which is natural in supersymmetry theories.

The rotation of $P$ may create ALP fluctuations by parametric resonance~\cite{Dolgov:1989us,Traschen:1990sw,Kofman:1994rk,Kofman:1997yn}, and the fluctuations may contribute to dark matter~\cite{Co:2017mop,Harigaya:2019qnl,Co:2020dya} with an abundance similar to or larger than the abundance given by kinetic misalignment, for $\epsilon = \mathcal{O}(1)$ or $\epsilon \ll 1$, respectively. For the latter case the prediction for $f_a$ becomes even smaller. The produced ALPs, however, tend to have a large velocities
\begin{align}
    v_a(T) \simeq 10^{-4} \left( \frac{T}{\rm eV} \right) \left( \frac{m_S}{0.1~{\rm GeV}} \right)^{1/2} \left( \frac{10^{-8}~{\rm eV}}{m_a} \right).
\end{align}
The warmness constraint~\cite{Viel:2013fqw} requires that $v_a({\rm eV})\lesssim 10^{-4}$ and restricts the model. This can be avoided if the produced ALPs are thermalized, or the rotation is close to circular motion and parametric resonance is absent.

\section{Summary and Discussion}
\label{sec:summary}

We discussed the possibility that an ALP has a non-zero velocity in the early universe and coupling with SM particles so that the baryon asymmetry of the universe is produced by electroweak sphaleron processes at the weak scale. The non-zero velocity of the ALP delays the beginning of the oscillation of the ALP around the minimum of the potential, and enhances the ALP abundance in comparison with the conventional misalignment mechanism.

From the requirement of simultaneously producing the observed baryon asymmetry and dark matter density, we obtain a prediction for the decay constant of the ALP, shown in Eq.~(\ref{eq:fa_prediction}). The corresponding predictions for the ALP-photon, -nucleon, and -electron couplings are summarized in Figs.~\ref{fig:photon}, \ref{fig:nucleon}, and \ref{fig:electron}, respectively. The predicted couplings are much larger than those of the QCD axion and of ALP dark matter from the conventional misalignment mechanism.
The predicted couplings can be probed by various experiments.

We assumed that the ALP explains the dark matter density. If we only require that the ALP velocity explains the baryon density, the predictions for the ALP couplings can be understood as lower bounds so that the ALP velocity does not overproduce ALP dark matter by kinetic misalignment.

\vspace{0.6cm}
\noindent
{\it Note added.} While finalizing the manuscript, Ref.~\cite{Domcke:2020kcp} appeared on arXiv, which also discusses the baryon asymmetry from general couplings of the ALP with standard model particles, and derives the dependence of the baryon asymmetry on the couplings. The paper focuses on the formulation of the computation of the coefficient $c_B$ relevant for ALP-genesis and does not discuss the prediction for the ALP couplings through ALP cogenesis.

\section*{Acknowledgment}
We thank Prateek Agrawal for useful discussions, and Jonathan L.~Ouellet for useful discussions and for providing the sensitivity curves for DM Radio. The work was supported in part by the DoE Early Career Grant DE-SC0019225 (R.C.), the DoE grants DE-AC02-05CH11231 (L.H.) and DE-SC0009988 (K.H.), the NSF grant NSF-1638509 (L.H.), as well as the Raymond and Beverly Sackler Foundation Fund (K.H.).

\appendix
\section{Estimation of baryon asymmetry}

In this appendix we estimate the coefficient $c_B$ in Eq.~(\ref{eq:nB}) for the standard model with left-handed quarks $q_i$, right-handed up quarks $\bar{u}_i$, right-handed down quarks $\bar{d}_i$, left-handed leptons $\ell_i$, right-handed electrons $\bar{e}_i$, and Higgs $H$. Here $i=1-3$ is the generation index. 

The Yukawa interactions are
\begin{align}
    {\cal L} = y^u_{ij}q_i \bar{u}_j H^\dag + y^d_{ij}q_i \bar{d}_j H + y^e_{ij}\ell_i \bar{e}_j H.
\end{align}
The couplings between the ALP $a = \theta f_a$ and the SM particles are
\begin{align}
    {\cal L} = \partial_\mu \theta \sum_{f,i,j}c_{f_{ij}} f^\dag \bar{\sigma}^\mu f +
    \frac{\theta}{64\pi^2}\epsilon^{\mu\nu\rho\sigma}\left(c_Y g'^{2} B_{\mu \nu} B_{\rho\sigma}+ c_W g^2 W^a_{\mu \nu} W^a_{\rho\sigma} + c_g g_s^2 G^a_{\mu \nu} G^a_{\rho\sigma} \right).
\end{align}
By unitary rotations, we can take $c_{f_{ij}} = \delta_{ij} c_{f_i}$.
The $U(1)_P$ charge density in the ALP, $\dot{\theta}f_a^2$, is transferred into the particle-antiparticle asymmetries of SM particles through the couplings between the ALP and the SM particles.
The Boltzmann equations governing the charge asymmetries are 
\begin{align}
\dot{n}_{q_i} = & 
\sum_j \gamma^u_{ij} \left( - \frac{n_{q_i}}{6} - \frac{n_{\bar{u}_j}}{3} + \frac{n_H}{4} + \frac{c_{q_i} + c_{\bar{u}_j}}{6}\dot{\theta} T^2 \right) 
+ \sum_j \gamma^d_{ij} \left( - \frac{n_{q_i}}{6} - \frac{n_{\bar{d}_j}}{3} -  \frac{n_H}{4} +  \frac{c_{q_i} + c_{\bar{d}_j}}{6}\dot{\theta} T^2\right)  \nonumber \\
& +  2 \Gamma_{\rm ss} \sum_j\left( -   n_{q_j} -n_{\bar{u}_j}-n_{\bar{d}_j} + \frac{2c_{q_j} + c_{\bar{u}_i} + c_{\bar{d}_j} - c_g/N_g}{2} \dot{\theta} T^2 \right),  \nonumber \\
& +  3 \Gamma_{\rm ws} \sum_j\Big(-n_{q_j} - n_{\ell_j} +  \frac{3c_{q_j} + c_{\ell_j}-c_W/N_g}{3} \dot{\theta} T^2  \Big)  \\
\dot{n}_{\bar{u}_i} = & 
\sum_j \gamma^u_{ji} \left( - \frac{n_{q_j}}{6} - \frac{n_{\bar{u}_i}}{3} + \frac{n_H}{4} + \frac{c_{q_j} + c_{\bar{u}_i}}{6}\dot{\theta} T^2 \right) \nonumber \nonumber \\
& + \Gamma_{\rm ss} \sum_j\left( -   n_{q_j} -n_{\bar{u}_j}-n_{\bar{d}_j} + \frac{2c_{q_j} + c_{\bar{u}_i} + c_{\bar{d}_j} - c_g/N_g}{2} \dot{\theta} T^2 \right), \\
\dot{n}_{\bar{d}_i} = & 
\sum_j \gamma^d_{ji} \left( - \frac{n_{q_j}}{6} - \frac{n_{\bar{d}_i}}{3} -  \frac{n_H}{4} +  \frac{c_{q_j} + c_{\bar{d}_i}}{6}\dot{\theta} T^2\right)\nonumber \\
&+ \Gamma_{\rm ss} \sum_j\left( -   n_{q_j} -n_{\bar{u}_j}-n_{\bar{d}_j} + \frac{2c_{q_j} + c_{\bar{u}_i} + c_{\bar{d}_j} - c_g/N_g}{2} \dot{\theta} T^2 \right), \\
\dot{n}_{\ell_i} = & \sum_j \gamma^e_{ij} \left( - \frac{n_{\ell_i}}{2} - n_{\bar{e}_j} -  \frac{n_H}{4}+ \frac{c_{\ell_i} + c_{\bar{e}_j}}{6}\dot{\theta} T^2 \right) \nonumber \\ 
& +  \Gamma_{\rm ws} \sum_j\Big(-n_{q_j} - n_{\ell_j} +  \frac{3c_{q_j} + c_{\ell_j}-c_W/N_g}{3} \dot{\theta} T^2  \Big) ,  \\
\dot{n}_{\bar{e}_i} = & \sum_j  \gamma^e_{ji} \left( - \frac{n_{\ell_j}}{2} - n_{\bar{e}_i} -  \frac{n_H}{4}+ \frac{c_{\ell_j} + c_{\bar{e}_i}}{6}\dot{\theta} T^2 \right), \\
\dot{n}_H = & \, -\sum_{ij}\gamma^u_{ij} \left( - \frac{n_{q_i}}{6} - \frac{n_{\bar{u}_j}}{3} + \frac{n_H}{4} + \frac{c_{q_i} + c_{\bar{u}_j}}{6}\dot{\theta} T^2 \right) + \sum_{ij}\gamma^d_{ij} \left( - \frac{n_{q_i}}{6} - \frac{n_{\bar{d}_j}}{3} -  \frac{n_H}{4} +  \frac{c_{q_i} + c_{\bar{d}_j}}{6}\dot{\theta} T^2\right) \nonumber \\
& + \sum_{ij}\gamma^e_{ij}  \left( - \frac{n_{\ell_i}}{2} - n_{\bar{e}_j} -  \frac{n_H}{4}+ \frac{c_{\ell_i} + c_{\bar{e}_j}}{6}\dot{\theta} T^2 \right),  \\
\dot{n}_\theta = & \, -\sum_{ij}(c_{q_i} + c_{\bar{u}_j})\gamma^u_{ij} \left( - \frac{n_{q_i}}{6} - \frac{n_{\bar{u}_j}}{3} + \frac{n_H}{4} + \frac{c_{q_i} + c_{\bar{u}_j}}{6}\dot{\theta} T^2 \right) \nonumber \\
& -\sum_{ij}(c_{q_i} + c_{\bar{d}_j})\gamma^d_{ij} \left( - \frac{n_{q_i}}{6} - \frac{n_{\bar{d}_j}}{3} - \frac{n_H}{4} + \frac{c_{q_i} + c_{\bar{d}_j}}{6}\dot{\theta} T^2 \right) \nonumber\\
&- \sum_{ij}(c_{\ell_i} + c_{\bar{e}_j})\gamma^e_{ij}  \left( - \frac{n_{\ell_i}}{2} - n_{\bar{e}_j} -  \frac{n_H}{4}+ \frac{c_{\ell_i} + c_{\bar{e}_j}}{6}\dot{\theta} T^2 \right)\nonumber \\
& -\sum_{ij}(2c_{q_i} + c_{\bar{u}_i} + c_{\bar{d}_i} - c_g/N_g)\Gamma_{\rm ss} \left( -   n_{q_j} -n_{\bar{u}_j}-n_{\bar{d}_j} + \frac{2c_{q_j} + c_{\bar{u}_i} + c_{\bar{d}_j} - c_g/N_g}{2} \dot{\theta} T^2 \right), \nonumber \\
&- \sum_{ij}(3c_{q_i} + c_{\ell_i}-c_W/N_g)\Gamma_{\rm ws} \Big(-n_{q_j} - n_{\ell_j} +  \frac{3c_{q_j} + c_{\ell_j}-c_W/N_g}{3} \dot{\theta} T^2  \Big), 
\end{align}
where
\begin{align}
    \gamma^u_{ij} \simeq \alpha_3 |y_{ij}^{u}|^2 T,~~
    \gamma^d_{ij} \simeq \alpha_3 |y_{ij}^{d}|^2 T,~~
    \gamma^e_{ij} \simeq \alpha_2 |y_{ij}^{e}|^2 T,~~
    \Gamma_{\rm ws} \simeq 10 \alpha_2^5 T,~~
    \Gamma_{\rm ss} \simeq 100 \alpha_3^5 T.
\end{align}
Here the dependence on $\dot{\theta}$ is derived in the following way~\cite{Co:2019wyp}. We consider a charge transfer from $\dot{\theta} f_a^2$ in each process, derive the would-be equilibrium values of the particle asymmetries via the process by minimizing the free-energy including the energy of the ALP, and use the principle of detailed balance.

The equilibrium values of asymmetries including all interactions are obtained by solving the equations $\dot{n}_f= \dot{n}_H=0$, with the conservation laws $Y = 0$ as well as $B/3-L_i=0$ if $y_{ij}^e$ is diagonal and $B-L=0$ otherwise. We find that
\begin{align}
    c_B =& \left( \frac{21}{158} - \delta \right) c_g - \frac{12}{79} c_W+ \sum_i\left( \frac{18}{79} c_{q_i} - \frac{21}{158}c_{\bar{u}_i} -  \frac{15}{158}c_{\bar{d}_i} + \frac{25}{237} c_{\ell_i}- \frac{11}{237}c_{\bar{e}_i} \right) \\
     =& \left( \frac{21}{158} - \delta \right) c_g - \frac{12}{79} c_W+ \sum_i\left( \frac{18}{79} c_{q_{ii}} - \frac{21}{158}c_{\bar{u}_{ii}} -  \frac{15}{158}c_{\bar{d}_{ii}} + \frac{25}{237} c_{\ell_{ii}}- \frac{11}{237}c_{\bar{e}_{ii}} \right) \nonumber
\end{align}
where in the second line we cast the formula into a basis-independent form. We define $\delta \simeq 0.005 \left( y_u/10^{-5}  \right)^2$ with $y_u$ the up quark Yukawa coupling.

Except for the coefficient of $c_g$, the coefficients can be derived by simply taking each term in the Boltzmann equation to be zero. This is because for $c_g=0$, a linear combination of the shift symmetry and fermion numbers remains exact. At the equilibrium, $\dot{n}_\theta$ should also vanish, and the whole system is in thermal equilibrium. We can apply the standard requirement that each term in the Boltzmann equation vanishes. For $c_g \neq 0$, since the shift symmetry is broken by the QCD anomaly and the quark Yukawa interaction, this argument is not applicable. One must use the whole Boltzmann equation to obtain the equilibrium values of the asymmetries of SM particles, for which $\dot{n}_\theta$ is non-zero; the system is not truly in equilibrium. In the limit where the up quark Yukawa vanishes, a symmetry becomes exact and we can use the standard argument and obtain the coefficient $21/158$ in front of $c_g$.

Note that the result is invariant under fermion field rotations that leave the Yukawa interactions invariant,
\begin{align}
    \begin{cases}
    \text{L rotations} &: c_{\ell_i} \rightarrow c_{\ell_i} + \alpha,~c_{\bar{e}_i} \rightarrow c_{\bar{e}_i} - \alpha,~c_W\rightarrow c_W + 3\alpha \\
    \text{B rotations} & : c_{q_i} \rightarrow c_{q_i} + \frac{\alpha}{3},~c_{\bar{u}_i,\bar{d}_i} \rightarrow c_{\bar{u}_i,\bar{d}_i} - \frac{\alpha}{3},~c_W\rightarrow c_W + 3 \alpha
    \end{cases}.
\end{align}

\newpage

\bibliography{ALP}

\end{document}